\documentclass[twocolumn,aip,jcp,lengthcheck,showpacs]{revtex4-1}
\usepackage[]{amsmath}
\usepackage{amssymb}
\usepackage{mhchem}
\usepackage[separate-uncertainty=true]{siunitx}
\usepackage{graphicx}
\usepackage{bm}
\usepackage{times,psfrag,subfigure}
\usepackage[]{color}
\usepackage{syntonly}
\usepackage{lmodern}
\DeclareMathOperator{\erf}{erf}
\DeclareMathOperator{\arctanh}{arctanh}
\begin{document}

\title{Chemotaxis and auto-chemotaxis of self-propelling artificial droplet swimmers} 

\author{Chenyu Jin}
\author{Carsten Kr\"uger}
\author{Corinna C. Maa\ss}
\email{corinna.maass@ds.mpg.de}
\thanks{Corresponding author}
\affiliation{Max Planck Institute of Dynamics and Self-Organisation, Am Fa\ss berg 17, 37077 G\"ottingen, Germany}
\date{published on PNAS, doi: 10.1073/pnas.1619783114
}

\begin{abstract}
Chemotaxis and auto-chemotaxis play an important role in many essential biological processes. 
We present a self-propelling artificial swimmer system which exhibits chemotaxis as well as negative auto-chemotaxis. 
Oil droplets in an aqueous surfactant solution are driven by interfacial Marangoni flows induced by micellar solubilization of the oil phase.
We demonstrate that chemotaxis along micellar surfactant gradients can guide these swimmers through a microfluidic maze. 
Similarly, a depletion of empty micelles in the wake of a droplet swimmer causes negative auto-chemotaxis and thereby trail avoidance.
We have studied auto-chemotaxis quantitatively in a microfluidic device of  bifurcating channels: branch choices of consecutive swimmers are anticorrelated, an effect decaying over time due to trail dispersion. We have modeled this process by a simple one-dimensional diffusion process and stochastic Langevin dynamics. Our results are consistent with a linear surfactant gradient force and diffusion constants appropriate for micellar diffusion, and provide a measure of auto-chemotactic feedback strength versus stochastic forces. This assay is readily adaptable for quantitative studies of both artificial and biological auto-chemotactic systems.
\end{abstract}

\maketitle
\section*{Introduction}
Locomotion of living bacteria or cells can be random or oriented.
Oriented motion comprises the various ``taxis'' strategies by which bacteria or cells react to changes in their environment \cite{brock}. 
Among these, chemotaxis is one of the best studied examples \cite{adler1966chemotaxis, hazelbauer2012bacterial}: Cells and microorganisms are able to sense certain chemicals (chemoattractants or chemorepellents), and move towards or away from them.
This is an essential function in many biological processes, e.g., wound healing, fertilization, pathogenic species invading a host or colonization dynamics \cite{cejkova2016chemo,wadhams2004making}. 
When the chemoattractant or chemorepellent is produced by the microorganisms themselves, the system exhibits positive or negative auto-chemotaxis. 
Thus, chemotaxis provides a mechanism of inter-individual communication. Modeling such communication strategies is key to understanding the collective behavior of microorganisms \cite{bonner1947evidence, zhao2013psl,drescher2010fidelity} as well as flocks of animals like fire ants \cite{couzin2009collective,jackson2006longevity}.

To model the swimming motion of microorganisms, various self-propelling artificial swimmer systems have been developed based on different mechanisms. 
Generally, there are two classes of swimmers: systems driven by and aligning with external fields \cite{lagzi2010maze, cejkova2014dynamics, abecassis2008boosting, dreyfus2005}, including chemotactic gradients, and self-propelled swimmers, which move autonomously in homogeneous environments \cite{maass2016swimming,toyota2009self,izri2014self,ismagilov2002autonomous,golestanian2005propulsion, howse2007self,buttinoni2012active}. 
Many autonomous swimmers additionally react to external fields, e.g., phototactic gradients~\cite{lozano2016phototaxis}.


Biological auto-chemotactic systems exhibit very complex behaviors \cite{budrene1995dynamics, brenner1998physical}, where physical effects are intermingling with effects from various bio-processes such as cell migration, metabolism and division.
To untangle these effects, there have been some design proposals for artificial systems such as in \cite{yashin2015designing}, and simulations on the dynamics of simple auto-chemotactic microswimmers \cite{tsori2004self, grima2005strong, sengupta2009dynamics, taktikos2011modeling, kranz2015effective}. 
Studies exist on collective effects like auto-chemotaxis induced clustering \cite{theurkauff2012dynamic,pohl2014dynamic,liebchen2015clustering},  but generally, there is still a lack of well-controllable and quantifiable experimental realizations of auto-chemotactic artificial swimmers.
We demonstrate chemotaxis and auto-chemotaxis in microfluidic geometries for a highly symmetric and tunable artificial model swimmer system: self-propelling oil droplets in an aqueous surfactant solution \cite{peddireddy2012solubilization,herminghaus2014interfacial, maass2016swimming}.

The quantitative study of chemotaxis  with traditional methods such as micropipette assays has been limited to observational studies \cite{berg1990chemotaxis,van2007biased}, as experimental conditions such as gradient strength are difficult to set in such geometries.
Using microfluidic techniques, experimental conditions can be much better controlled, e.g., a linear gradient can be generated and kept constant, or even fast switched \cite{jeon2002neutrophil,irimia2006microfluidic}, the object distribution can be easily analyzed \cite{mao2003sensitive}, and the objects can be tracked individually \cite{song2006dictyostelium, amselem2012control,amselem2012stochastic}.
In this paper we present a microfluidic assay for the quantitative study of auto-chemotaxis.
We have not only observed auto-chemotaxis reproducibly, but also been able to directly measure system parameters like diffusion constants. 
This enables further quantitative experimental studies on the dynamics of simple auto-chemotactic swimmers.

\section*{Self-propelling droplet swimmers}
\begin{figure}
  \centering
  \includegraphics[width=\columnwidth]{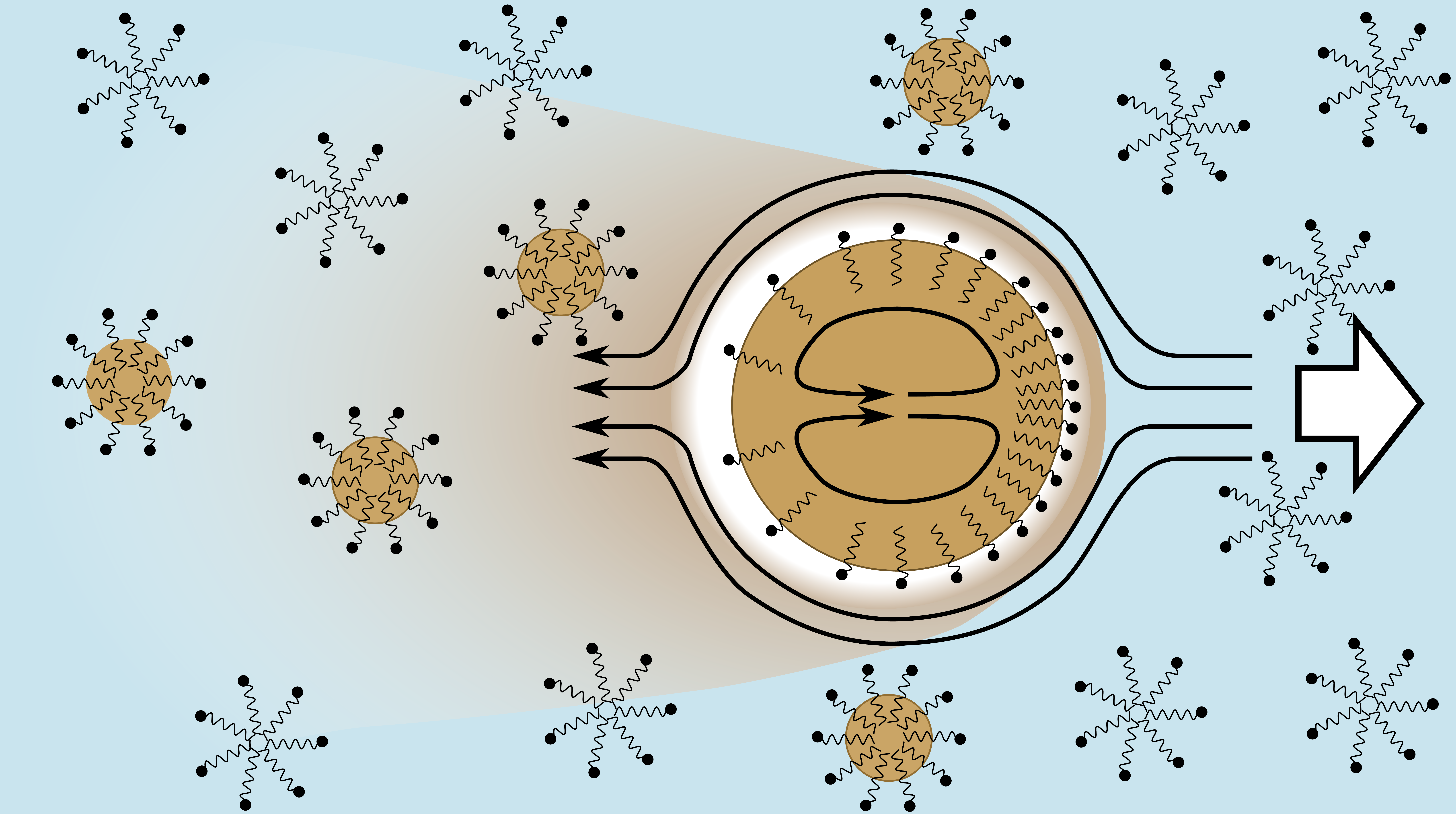}
  \caption{Schematic drawing of a droplet swimmer moving in surfactant solution. Due to the micellar solubilization of the oil phase, filled micelles disperse from the droplet into the solution, as shown by the yellowish shadow.
  Close to the droplet the free surfactant molecules are depleted, as shown by the white circle.
  When the droplet moves, it encounters more empty micelles in the front and leaves more filled micelles behind. 
  The inhomogeneous interfacial surfactant coverage on the droplet then starts the Marangoni flow and sustains the motion. Drawing not to scale. 
  }
  \label{fig:scheme}
\end{figure}

When an oil droplet dissolves in a surfactant solution, oil molecules will continuously migrate into the surfactant micelles until the entire droplet is solubilized. 
The final equilibrium state of the system is a homogeneous micellar nanoemulsion, comprised of a mixture of empty micelles, oil-filled micelles, and free surfactant molecules at the critical micelle concentration (CMC). The droplet locomotion is caused by a self-sustained Marangoni flow due to the inhomogeneous interfacial surfactant coverage and only observable in the non-equilibrium state of solubilization (Fig.~\ref{fig:scheme}) \cite{herminghaus2014interfacial}. While incorporating oil molecules, micelles grow in size, incorporating free surfactant molecules from the aqueous phase and increasing the total area of oil-water interfaces in the system.
A boundary layer forms around the droplet with a reduced density of free surfactant which in turn depletes the surfactant coverage of the droplet interface. 
This depletion is counteracted by the disintegration of empty micelles either approaching the droplet via diffusion or, if the droplet is moving, via advection. 
Advection will lead to more available empty micelles in front of the droplet and a trail of filled micelles behind it. In consequence, the depletion at the droplet apex is less pronounced and the resulting Marangoni flow will drive the droplet further forward towards even more empty micelles. 
At sufficiently high surfactant concentrations, small fluctuations in the droplet position or surfactant density are sufficient to start sustained self-propulsion. 
In flow equilibrium, set by the balance of Marangoni forces and viscous dissipation, the swimmer moves at a constant speed controlled by the global surfactant concentration.

The system presented in this study uses the ionic surfactant tetradecyltrimethylammonium bromide (TTAB); the oil phase consists of the nematic liquid crystal 4-pentyl-4'-cyano-biphenyl (5CB).  Nematic droplet swimmers of pure 5CB exhibit a strong curling instability~\cite{kruger2016curling} in their propulsion, which is absent in isotropic droplets. For this study, we only use isotropic droplets, either by keeping the ambient temperature above the nematic-isotropic transition at $T_\text{NI}=\,$\SI{35}{\celsius}, or by substituting a mixture of 5CB and 1-Bromopentadecane (BPD) with a volume ratio of 10:1. Henceforth, we will refer to the respective droplet types using the notations 5CB and 5CB/BPD.
Droplets are mass produced in microfluidic flow-focusing devices with high monodispersity ($<5 \%$) and sizes adjustable between 30 and 100 \si{\um}. 
The observation time is on the order of hours. Droplet speed and trajectory persistence are well controllable via temperature and surfactant concentration \cite{kruger2016curling}. 
The droplet propulsion is initiated at surfactant concentrations above $4\,\text{wt} \%$ TTAB.

\section*{Chemotactic maze solving}
Following the argument above, an external gradient of surfactant, i.e. empty micelles, will result in an alignment of the Marangoni flow with the gradient direction; as a result the droplet swimmer will move towards higher surfactant concentrations.
This behavior is typical for chemotaxis: the swimmer has no preferred direction in a homogeneous medium, yet moves directionally in the presence of a chemical gradient. 

\begin{figure}
  \centering
  \includegraphics[width=\columnwidth]{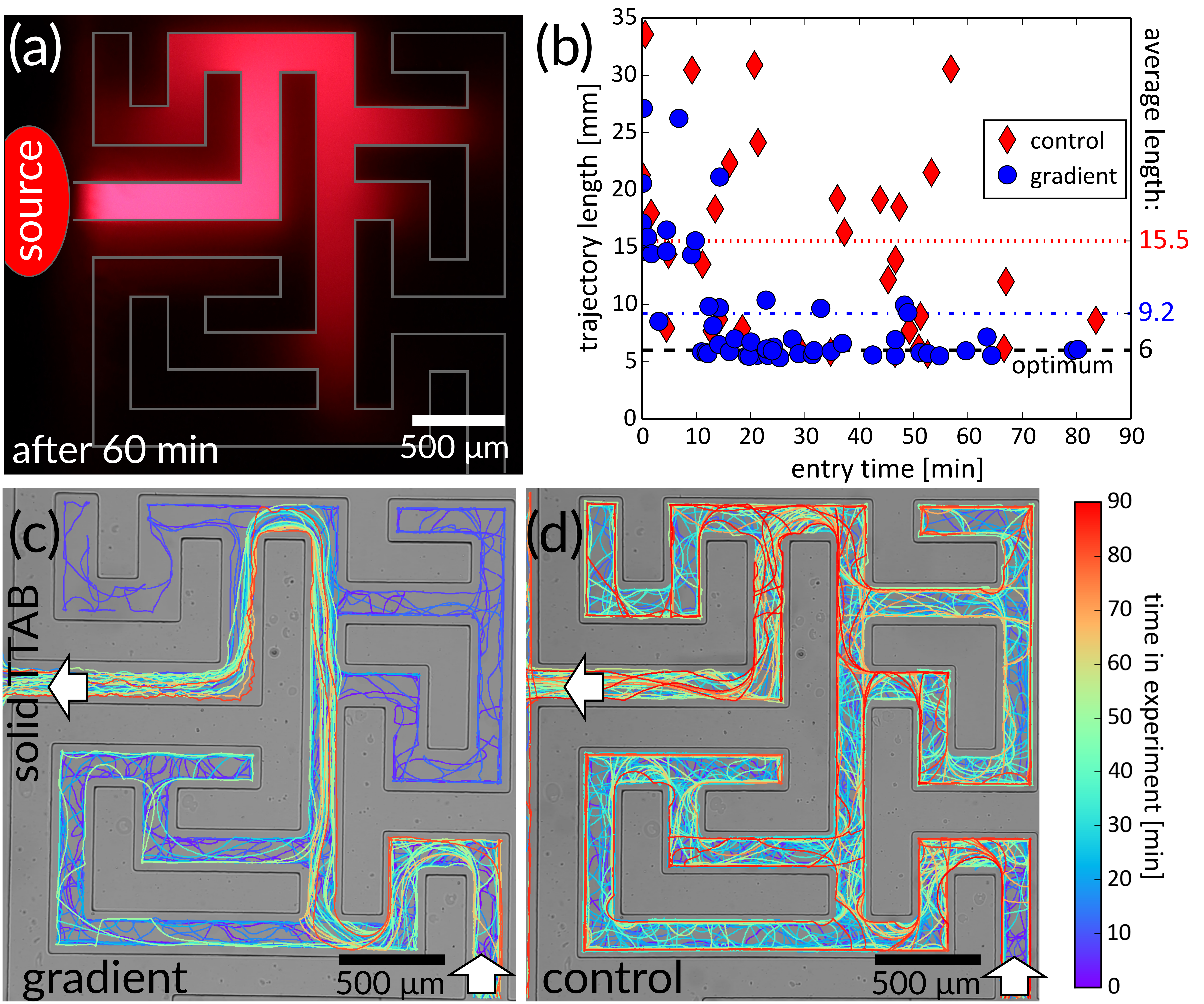}
  \caption{Maze solving by chemotactic droplet swimmers. White arrows indicate maze entrance and exit. 
  (a) Solid TTAB mixed with fluorescent Nile Red spreading in the maze; distribution after \SI{60}{\minute}. ``Source'' marks the point of release (the  excitation LED was shaded in this area to improve contrast).
  (c) and (d) show trajectories with and without TTAB gradient; we selected only swimmers that passed both entrance and exit points. Line colors correspond to the time in the experiment. In (c), detours are mostly for early times (purple) while in (d) there is no correlation.  (b) Plot of path lengths vs. entry time, compared to the shortest path length ($6$ mm). See also SI movies 1,2, and 3.}
  \label{fig:maze}
\end{figure}

To demonstrate the chemotactic nature of our droplet swimmers, we used a design inspired by Lagzi et al. \cite{lagzi2010maze} consisting of two reservoirs connected by a microfluidic maze (Fig.~\ref{fig:maze}). Chemoattractant released at the exit spreads into the maze,  with the local concentration depending on the path distance to the exit. By moving up gradients, swimmers will then prefer the shortest path, as shown in chemotactic experimental systems \cite{lagzi2010maze, cejkova2014dynamics} and simulations \cite{kim2008investigations}. 

To initiate droplet propulsion, the maze is prefilled with a micellar TTAB solution at $5\,\text{wt} \%$.
Directly after the droplet swimmers (5CB/BPD) are released into the entrance reservoir, solid TTAB is added to the exit reservoir, acting as a chemoattractant. The TTAB gradually dissolves and spreads into the maze via convective diffusion, i.e. significantly faster than simple micellar diffusion. 
Thus, there will be a positive gradient along the optimum path through the maze, attracting the swimmers, whereas dead ends and detours will feature negative gradients, repelling the swimmers back towards the shortest path.
In a control experiment, we prefilled the maze again with a $5\,\text{wt} \%$ TTAB solution, but added no solids, such that the overall concentration was homogeneous.

Fig.~\ref{fig:maze} shows results from our experiment. In panel (a), we imaged the surfactant spreading inside the maze qualitatively by mixing the solid TTAB with the fluorescent Nile Red dye, which is insoluble in water and therefore co-moves with the surfactant micelles. 
The still image in panel (a) is taken \SI{60}{\minute} after the release of solid TTAB; the additional surfactant has spread to the maze entrance and its concentration decreases along side branches.

Fig.~\ref{fig:maze}(c) and (d) show the trajectories of swimmers in a gradient (c) and control experiment (d) during a \SI{90}{\minute} time interval. The trajectories are color-coded by time from blue to red. 
In the presence of a surfactant gradient, (c), the trajectory density is highest along the shortest path, with detours occurring primarily in the first \SI{20}{\minute}, before the surfactant has spread sufficiently. In the control experiment without a gradient, (d), the swimmers move freely and explore the entire maze, with no correlations in time.

Panel (b) compares trajectory lengths between gradient and control experiments for all swimmers that successfully traversed the maze, sorted by the time at which they entered the maze. 
Initially, trajectory lengths are comparable between both experiments. After \SI{20}{\minute}, in the presence of a well established gradient (blue circles), over 80\% of the recorded trajectories approach the optimum length of \SI{6}{\mm}, with an average trajectory length of \SI{9.2}{\mm} over the entire duration of the experiment. Without a gradient (red diamonds), the trajectory length is on average (\SI{15.5}{\mm}) more than twice the optimum length and there is no time dependence. Time lapse movies for panels (a), (c), and (d) can be found in the supporting information (movies 1--3).

\section*{Swimmers exhibiting negative auto-chemotaxis}
\begin{figure}
  \centering
  \includegraphics[width=\columnwidth]{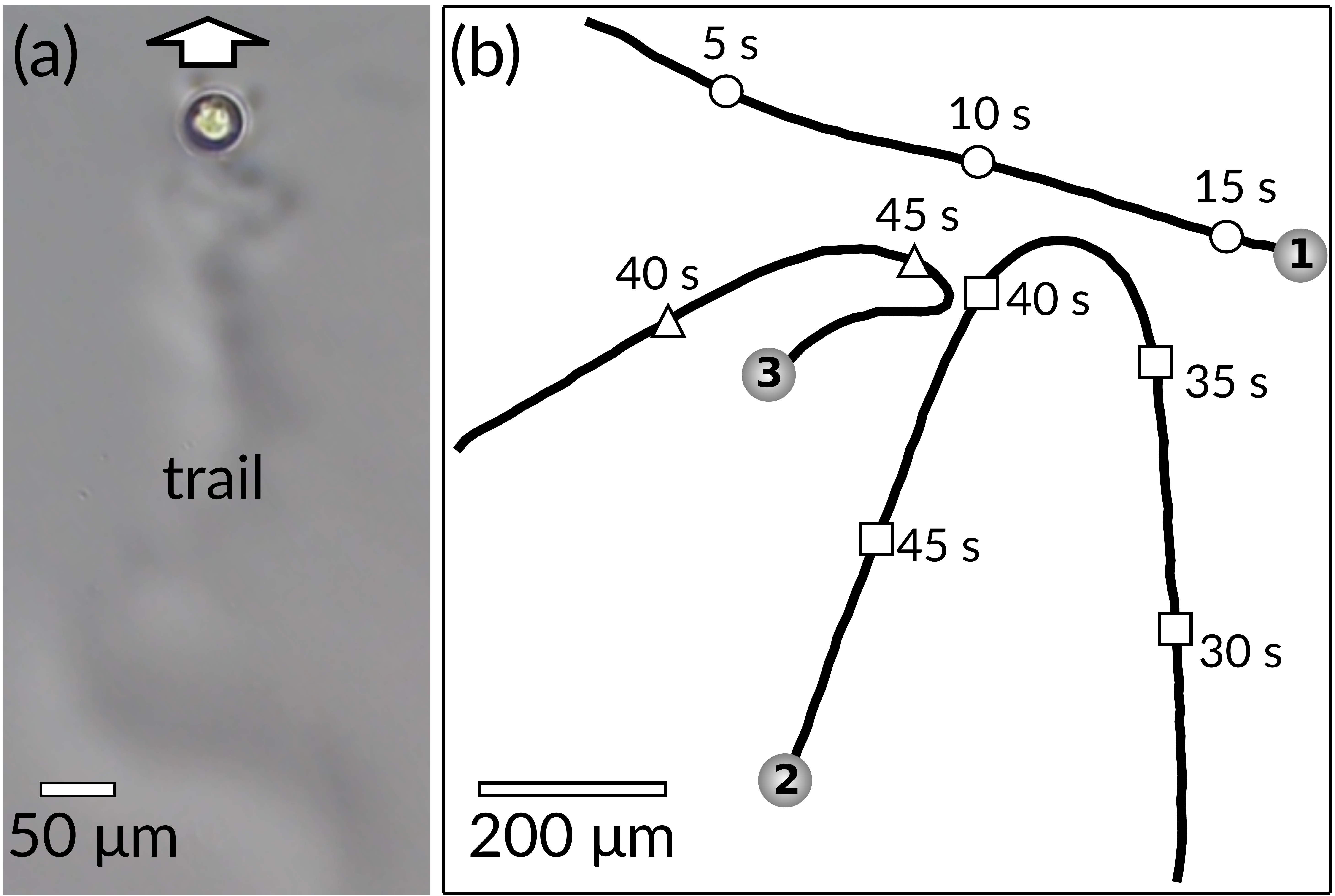}
  \caption{(a) A droplet swimmer leaves a trail which can be seen under phase contrast microscopy from the slightly different refractive index. (b) Free swimmers avoiding each other's trails. Sphere drawings (to scale) mark the trajectory end points, timing marks on trajectories time points in the experiment. See also SI movies 4 and 5.}
  \label{fig:freeswimmer}
\end{figure}

While a droplet swimmer moves through the solution, it will leave a trail of filled micelles behind, such that the fraction of empty micelles in the trail decreases.
Since the swimmers are sensitive to the density of empty micelles, they will therefore avoid their own trail, i.e. exhibit negative auto-chemotaxis.

This is shown qualitatively in Fig.~\ref{fig:freeswimmer}. Panel (a) shows a 5CB droplet swimming  in a surfactant solution of high concentration ($25\,\text{wt}\%$ TTAB) and high temperature (\SI{37}{\celsius}), resulting in a high solubilization rate. A trail of oil filled-micelles, dispersing over time in the wake of the droplet, can be inferred from variations in the refractive index in the aqueous phase.

In panel (b) we have plotted three trajectories of isotropic 5CB swimmers ($T=$ \SI{37}{\celsius}, $7.5\,\text{wt}\%$ TTAB) interacting in a Hele-Shaw cell, with markers on the trajectories labeling the time since the start of observation.  
The first swimmer, marked \textbf{1}, moves in an unperturbed manner, but swimmer \textbf{2} is repelled from swimmer \textbf{1}'s trail, even though swimmer \textbf{1} passed \SI{20}{\second} earlier.  Swimmer \textbf{3} approaches swimmer \textbf{2}'s trajectory more closely, \SI{5}{\second} after \textbf{2}'s passage, but is turned away sharply. This is consistent with a steep gradient in filled micelles in the not yet strongly dispersed trail directly behind a swimmer, leading to a stronger and more short-ranged chemotactic repulsion than in the interaction between swimmers \textbf{1} and \textbf{2}. Note that due to the quasi 2D geometry of the cell, hydrodynamic interactions are suppressed for droplet distances exceeding the cell height of \SIrange{50}{55}{\micro\metre}.

\section*{Branch choice by auto-chemotactic signaling}
Auto-chemotactic processes are generally treated considering the following aspects: the secretion and decay of the chemical constituting the trail, the diffusion of the trail, and the interaction of the swimmer with the self-generated chemical gradient.
Appropriate simulations on auto-chemotaxis \cite{tsori2004self, sengupta2009dynamics} have been conducted using parameters extracted from experimental studies \cite{brenner1998physical, van2007biased}.
Inspired by Ambravaneswaran et al.~\cite{ambravaneswaran2010directional}, we have designed a microfluidic experiment to study auto-chemotactic processes quantitatively by having multiple swimmers consecutively pass a series of bifurcations in a channel. 

The auto-chemotaxis problem is reduced to a simple measure of correlated binary branch choices, resulting in a high statistical yield. We have fitted such correlation data with an analytical model balancing a gradient force against a stochastic noise term. In the model, we assume that the coupling between the gradient in filled micelles and the repulsive chemotactic force is linear and that micellar diffusion determines the gradient evolution over time.

\begin{figure}
  \centering
  \includegraphics[width=\columnwidth]{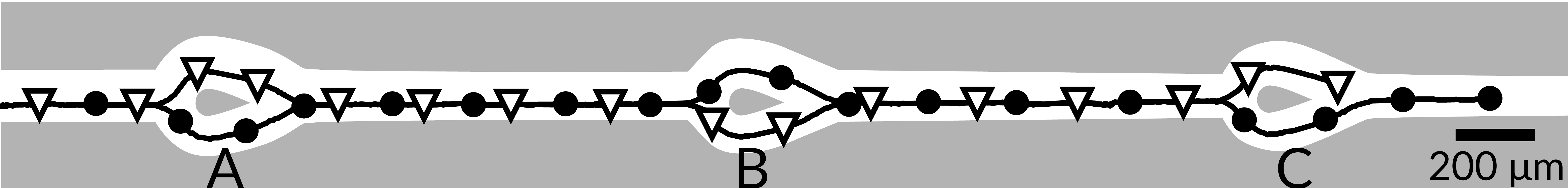}
  \caption{Swimmers moving through bifurcating channels choose alternating branches. Two trajectories of different swimmers are plotted in lines and marked by dot (first) and triangle (second) in the order of passage. See also SI movie 6 and Fig.~S1 (cell design).
  }
  \label{fig:branch}
\end{figure}

We begin with an example experimental run: in Fig.~\ref{fig:branch}, we have drawn a channel pattern with three bifurcations A, B and C (white mask) and overlaid it with two selected trajectories for 5CB/BPD swimmers. 
The channel connects an entrance with a large exit reservoir; symmetric bifurcations are generated by a series of pillars, which are tear-shaped with the pointed end facing the exit reservoir to keep the swimmers from turning back around the pillar. 
This pinch-off design is quite efficient: in our experiments, $75\%$ of swimmer interactions with the pillar contained only one passage through a single branch.
Note that there is no overall flow or surfactant gradient between the entrance and exit of the channel. 
The trajectories are marked by dot (swimmer \textbf{1}) and triangle (\textbf{2}), in the order of passage. Swimmer \textbf{2} enters the channel approximately \SI{20}{\second} after swimmer \textbf{1}.  The two swimmers are anti-correlated in their choice at all bifurcations. 

The trail secretion of a solubilizing droplet ($\beta$ molecules per swimmer per second) can be established from the solubilization rate - i.e., the time dependent droplet size. Contrary to most biological auto-chemotactic systems, there is no decay of the secretion products. 
Since the width and height of the channel, $w=$ \SI{100}{\um}, $h=$ \SI{110}{\um}, are of the order of the initial swimmer diameter $2r=\SI{100}{\um}$, 
we assume the filled micelle density to be constant over the channel cross section. We can therefore map our model of trail diffusion and auto-chemotactic branch choice to a one-dimensional (1D) problem along the channel midline around the pillar. 
The number density of the solubilized oil molecules in the trail directly behind the swimmers is $c_0 = \beta /(vwh)$, with the speed of the swimmers $v$ and the channel width $w$ extracted from the experiment.
We approximate the initial secretion profile $c(x)$ in the channel by a 1D step function. The average duration of a channel passage is ca.~\SI{15}{\second} and we set the time origin $t=0$ to the instant when the first swimmer leaves the bifurcation at $x=-l$.  Since the step function approximation is not valid for short times and we expect pressure equilibration flows around the pillar when the first droplet leaves, we only use events for data fitting where the second droplet enters the junction more than \SI{20}{\second} after the first has left ($\Delta t>$ \SI{20}{\second}).

\begin{figure}
  \centering
  \includegraphics[width=\columnwidth]{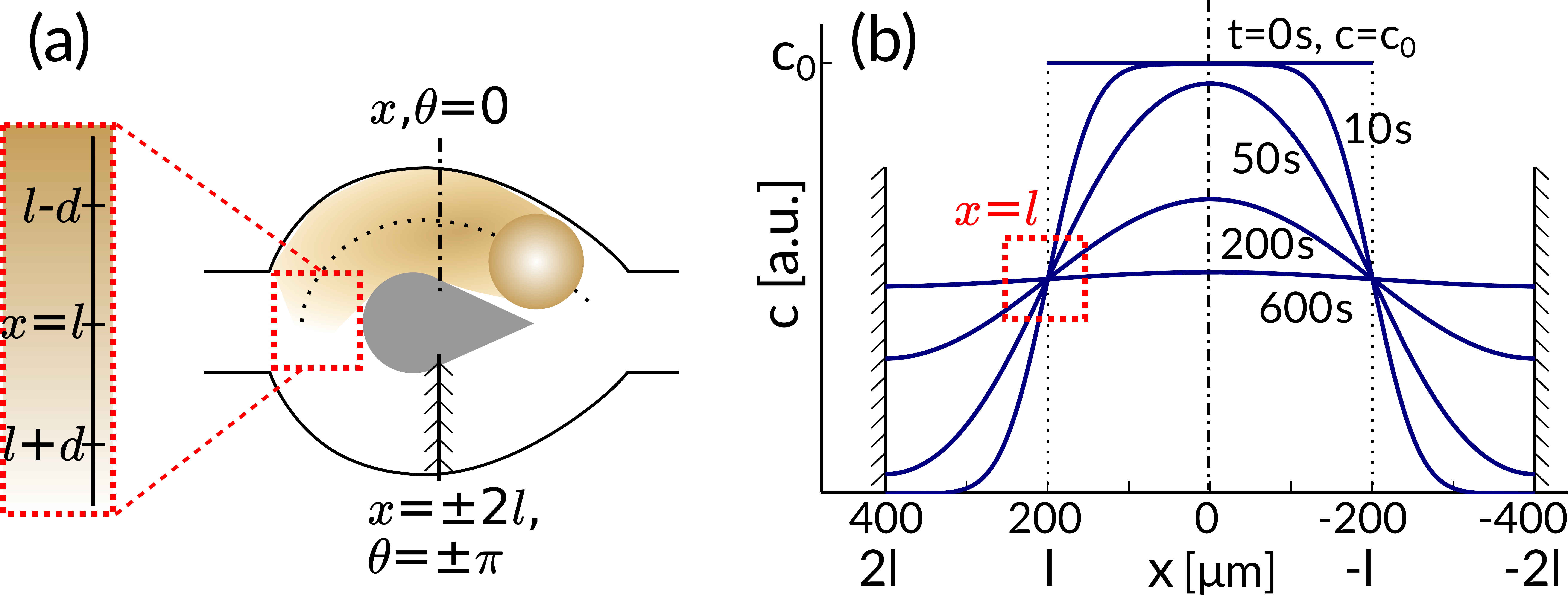}
  \caption{
    (a) Schematic drawing of a bifurcation. The trail of a swimmer fills the top branch and diffuses into the bottom branch around the pillar. The bifurcation is mapped to a fixed radius polar coordinate $x=(2l/\pi)\cdot\theta$ with the origin at the top of the bifurcation and $x=l$ at its entrance. (b) Trail dispersion, characterized by the concentration profile of micellar-solubilized oil molecules $c$, approximated by 1D diffusion from a step function between reflecting boundaries, with calculated profiles for times between \SI{0}{\second} and \SI{600}{\second}. Red dotted rectangles mark the gradient at $x=l$.  
  }
  \label{fig:branch-ana}
\end{figure}

To model the diffusion of the oil-filled micelles, we approximate the bifurcation by a circular pillar of radius $R=2l/\pi$ and use a polar coordinate at fixed radius $x=R\cdot\theta$ with the angle $\theta=0$ at the top of the pillar, progressing counterclockwise, such that the entrance bifurcation is at $x=l$ (see Fig.~\ref{fig:branch-ana} (a)). Without loss of generality, we assume that the swimmer enters the bifurcation from the left at $x=l$ and chooses the top branch. 

In our circular pillar approximation, the diffusion problem is symmetric around $x=0$, and we neglect diffusion into the main channel at $x=\pm l$. 
The gradient evolution at $x=l$ can be mapped onto the problem of 1D diffusion between two reflecting or periodic boundaries at $x=\pm 2l$ (Fig.~\ref{fig:branch-ana}(b)).
The concentration profile in the region $-2l \leq x \leq 2l$ evolving over time from a step function $c(-l \leq x \leq l) = c_0$ is \cite{crank1979mathematics}  

\begin{equation}
  c(x,t) =
   \frac{c_0}{2} \sum_{n=-\infty}^{\infty} \left(\erf \frac{(4n+1)l-x}{\sqrt{4D_\text{f}t}} - \erf \frac{(4n-1)l-x}{\sqrt{4D_\text{f}t}} \right),
  \label{eqn:c-erf}
\end{equation}
where $D_\text{f}$ denotes the diffusion coefficient of the filled micelles, which are the carriers of the solubilized oil. We have provided a full derivation in the supporting information (section S1).

Using parameters appropriate to our experimental system ($l \approx$ \SI{200}{\um}, $D_\text{f} \approx$ \SI{100}{\square\um\per\second}  \cite{candau1984new}), we have calculated and plotted examples of concentration profiles at different times in Fig.~\ref{fig:branch-ana}(b).
We expect the anti-correlation between the choices of two consecutive swimmers to decay in time, depending on the decrease of the gradient at $x=l$.
In the long time limit, after the environment in the two branches becomes homogeneous again, the choice of a swimmer between the two branches should be completely random, i.e., independent of the choice of the previous swimmer.

\begin{figure}
  \centering
  \includegraphics[width=\columnwidth]{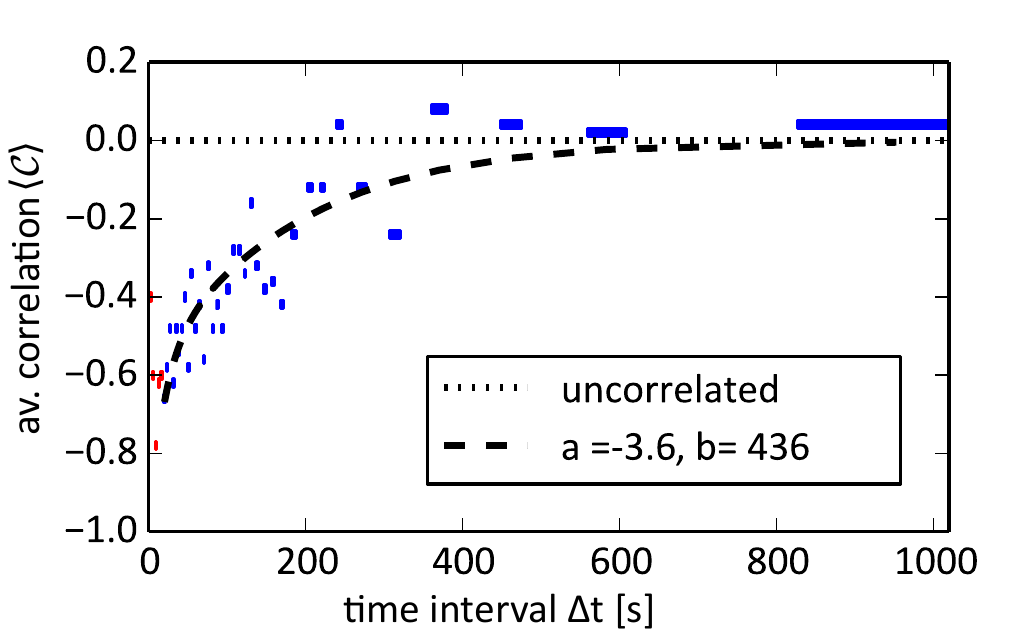}
  \caption{
        Correlation $\langle\mathcal{C}\rangle$ between branch choices of consecutive swimmers vs. the time interval $\Delta t$ between passages. Data were binned by $\Delta t$ using $100$ events/bin and then averaged (the corresponding  $\Delta t$ range is marked by the bar width). $\langle\mathcal{C}\rangle$ decorrelates with increasing $\Delta t$, with the limits of $\langle\mathcal{C}\rangle\in[-1,0]$ for perfect anticorrelation (-1) and no correlation (0). The parameters $a=-3.6\pm 0.2$ and $b=(4.3\pm 1.2)\times 10^2\si{\second}$ for the fitted $\langle\mathcal{C}\rangle=\tanh (\xi)$ were derived by fitting $\xi$ using Eqns.~\ref{eqn:Ac-tanh},\ref{eqn:Ac-fit} and \ref{eqn:ab}. Correlation data for  $\Delta t<\SI{10}{\second}$ (red bars) were omitted from the fitting to rule out hydrodynamic drag effects.}
  \label{fig:correlation}
\end{figure}

The swimmers make their choice at the entrance $x=l$ under both the gradient force and the stochastic force.
In all experiments, we have never observed swimmers reversing direction once past the bifurcation.
This indicates that the choice is made within a small region $x\in[l-d,l+d]$, as sketched in Fig.~\ref{fig:branch-ana} (a), with $d$ on the order of the droplet diameter or channel width, and on a timescale $\tau$ much shorter than the timescale $t$ of the trail dispersion. Hence, the motion of the swimmer $x(\tau)$ during the decision process can be approximated as a 1D Brownian motion under a constant gradient force between two absorbing boundaries at $l\pm d$, described by the following overdamped Langevin equation:
\begin{equation}
  \left.\frac{dx}{d\tau}\right|_{x=l} = \kappa \partial_x c + \sqrt{2D} \Gamma(\tau).
  \label{eqn:langevin}
\end{equation}
Here, $\kappa$ is a linear coefficient quantifying the sensitivity to the chemorepellent gradient $\partial_x c$. Since the chemorepellent is self-produced, $\kappa$  corresponds to the auto-chemotactic feedback strength used in literature \cite{tsori2004self, grima2005strong, sengupta2009dynamics, taktikos2011modeling, kranz2015effective}. Note that $\kappa$ is not identical to the chemotactic strength as applicable to the maze experiments, which is based on a gradient in empty micelles (see section S3 in the supporting information).
$\Gamma(\tau)$ is a normalized Gaussian noise term with $\langle \Gamma(\tau) \rangle =0$, $\langle \Gamma(\tau), \Gamma(\tau+\Delta \tau) \rangle = \delta(\Delta \tau)$. The velocity of the active swimmer is not included in Eq.~\eqref{eqn:langevin}, as the branch choice direction is orthogonal to the incoming swimmer. $D$ denotes the  diffusion coefficient of a passive swimmer and is presumably larger than the Stokes-Einstein value of $k_BT/6\pi\eta r$, since the micellar solubilization process provides an additional source of stochastic noise.

The position of the swimmer $x(\tau)$ can be written as
\begin{equation}
  x(\tau)-l = \kappa \partial_x c\,\tau + \sqrt{2D} B^\circ(\tau)
  \label{eqn:y}
\end{equation}
where $B^\circ(\tau)$ is a standard Brownian motion process and $x(0) =l$.

The probability $\mathcal{P}$ of anticorrelated branch choices between two consecutive swimmers is then the probability that the biased Brownian motion process $x(\tau)$ reaches $l+d$ before $l-d$ \cite{stochastic}:
\begin{align}
  \mathcal{P} &= \frac{ \displaystyle{ 1- \exp \left( -2\xi \right) }}{ \displaystyle{ \exp \left(2\xi \right) - \exp \left(-2\xi\right) }}, & \xi &= -\frac{\kappa \cdot d}{2D} \partial_x c. 
  \label{eqn:Pr}
\end{align}

In our data analysis, we record events of correlated branch choices between consecutive swimmers as $\mathcal{C}=1$ and anticorrelated choices as $\mathcal{C}=-1$. If the interaction of a swimmer and a pillar contains several passages, i.e. the swimmer orbits the pillar, we consider only the last passage of the preceding and the first passage of the following swimmer. 
To study the time dependent decay of the (anti-)correlation, we bin the experimental result according to the time interval $\Delta t$ between the preceding swimmer leaving the bifurcation and the following swimmer entering it.
Using \eqref{eqn:Pr} with a concentration gradient $\partial_x c$ at ${x=l,t=\Delta t}$, the average correlation between the choices of consecutive swimmers is

\begin{equation}
  \langle\mathcal{C}\rangle = -1 \cdot \mathcal{P} + 1 \cdot (1-\mathcal{P}) =  \tanh \left( \xi(\Delta t)\right). 
  \label{eqn:Ac-tanh}
\end{equation}

The statistical result for $\langle \mathcal{C} \rangle$ from a series of branch-choosing experiments is shown in Fig.~\ref{fig:correlation}. Since some microfluidic bifurcations can be biased due to fabrication errors, we only accepted results from bifurcations where the overall preference for a single branch was less than 75\%. 
The bias corrected data set contains 4160 correlation events, omitting 283 rejected events.
The data are binned by the time interval $\Delta t$; the average correlation $\langle \mathcal{C} \rangle$ of each bin is plotted versus the corresponding average $\Delta t$ (blue bars).
To account for the steep correlation decay for short times and decreasing statistics for long times, we use a constant number of events (100 swimmer pairs) per bin, resulting in an increasing range of time intervals, as indicated by the bar width.

When $\Delta t$ is small, the choices of swimmers show a clear anti-correlation up to $-0.8$, i.e., $90\%$ of the following swimmers choose the branch that the preceding swimmer did not pass.
As $\Delta t$ increases, this average anti-correlation decreases to values close to zero, i.e., swimmers enter branches randomly and independent of the preceding swimmer. 

With \eqref{eqn:Ac-tanh}, $\xi$ can be easily calculated from the anti-correlation data.
We truncate the concentration profile from \eqref{eqn:c-erf} to the $n= 0, \pm 1$ terms and fit $\xi$ with 
\begin{equation}
  \xi =- \frac{a}{\sqrt{t}}(2 \exp (-b/t) -1 - 2 \exp (-4b/t) + \exp (-9b/t) ),
  \label{eqn:Ac-fit}
\end{equation}
with two parameters, a prefactor $a$ and a time constant $b$:
\begin{align}
a &= \frac{\kappa}{D} \cdot\frac{\beta \,d}{4vwh \sqrt{\pi D_\text{f}}}, & b =\frac{l^2}{D_\text{f}}.
\label{eqn:ab}
\end{align}
The corresponding function for the average correlation $\langle \mathcal{C} \rangle$ is then plotted in Fig.~\ref{fig:correlation} as black dashes.

With $b \approx$ \SI{436}{\second} from fitting the experimental data and $l \approx$ \SI{200}{\um}, we calculated $D_\text{f} \approx$ \SI{92}{\square\um\per\second}, which agrees with calculated and literature values ($D_\text{f} \approx$\,\SI{100}{\square\um\per\second}) \cite{candau1984new}.
In our expression for $a$ in \eqref{eqn:ab}, all quantities except the auto-chemotactic coupling strength $\kappa$ and the droplet diffusion coefficient $D$ can be measured or calculated independently. We can therefore use $a$ as a direct measure of the strength of chemotactic vs. stochastic forces, $\kappa/D$.

\section*{Conclusion and Summary}
We have studied a system of self-propelled droplets which exhibits chemotaxis comparable to biological systems. 
The motion of the oil-in-water droplets is driven by a Marangoni flow, which is caused by a self-sustained interfacial tension gradient during the solubilization of the droplet in a micellar surfactant solution.
In a homogeneous solution, the swimmers show no directional preference, whereas they move up surfactant gradients, i.e., gradients in empty micelles, which effect we have used to guide them through microfluidic mazes.

A related effect is trail avoidance by negative auto-chemotaxis, due to gradients of oil filled micelles in the wake of a droplet.
We have studied this effect quantitatively, observing anticorrelated branch choices between consecutive droplets in microfluidic channel bifurcations.
We could model the time dependent correlation decay analytically, assuming a force on the droplets proportional to the local empty micelle gradient. 

Fitting our data yielded two system parameters:
a time constant $b$ depending on the micellar diffusion timescale and a linear prefactor $a$, containing the secretion rate $\beta$ and the strength of (auto-)chemotaxis $\kappa$ over stochastic force $D$.
In biological systems, where various physical effects and bio-processes are intertwined, these parameters are often difficult to measure independently.
Microfluidic assays as presented above can provide reproducibly quantitative experimental data for statistics and comparative studies.
In our droplet swimmer system, we will use insights from this study to predict and control auto-chemotactic effects in more complicated geometries and to compare the swimmer dynamics with theoretical models \cite{tsori2004self, grima2005strong, sengupta2009dynamics, taktikos2011modeling, kranz2015effective}.

\section*{Material and methods}
\subsection*{Chemicals}
We obtain 5CB, BPD, TTAB and Nile Red from commercial suppliers (Synthon Chemicals and Sigma Aldrich) and use them as is.

\subsection*{Microfluidic devices}
We fabricate microfluidic devices using standard soft lithography procedures:
We create photomasks in a 2D AutoCad application and have them printed as a high resolution emulsion film by an external company (128k DPI, JD Photo-Tools, UK).
A Si wafer (Wafer World Inc.) is spin coated with a negative photoresist (SU-8, Micro Resist Technology) in a clean room environment.
UV light exposure through a photomask and subsequent chemical development produce a master wafer containing the microstructures.

We then use the master wafer in a polymer molding step to cast the microstructure into PDMS (Sylgard 184, Dow Corning, USA).
After degassing and heat curing at \SI{75}{\degreeCelsius} for $2$ hours, we peel the PDMS replica off from the wafer, cut it into single pieces, and punch in fluid inlets and outlets.
We then seal the molded PDMS blocks from below by glass slides. Covalent bonding between PDMS and glass is achieved by pretreating all surfaces in an air plasma (Pico P100-8, Diener electronic GmbH + Co. KG, Germany) for \SI{30}{\second}. 

We produce droplets in standard flow-focusing microfluidic devices, mounting syringes (Braun) on a precision microfluidic pump (NEM-B101-02B, cetoni GmbH, Germany) and connecting them to the inlets and outlets with Teflon tubing (39241, Novodirect GmbH). 
To create oil-in-water emulsions, we first activate the originally hydrophobic PDMS surfaces by a 1:1 volumetric mixture of \ce{H2O2}/\ce{HCl}, then fill the channels with a silanization solution (\ce{(C2H2O)_n C7H18O4Si}) for \SI{30}{\minute} and finally rinse them with milli-Q water. 

During experiments, we introduce a concentrated droplet stock solution into microfluidic devices using standard pipettes.

\subsection*{Image recording and analysis}
We record swimmers (still images and video) on a Olympus IX-73 optical microscope connected to a commercial DSLR camera (Canon EOS 600D) at 4 frames per second. 
Images are processed (swimmer tracking and trajectory analysis) using software written in-house in Python/openCV. 

\section*{Acknowledgement}
 Discussion with R. Breier, F. Schwarzendahl, S.~Herminghaus and C.~Bahr are gratefully acknowledged. C.~Jin is supported by the BMBF/MPG ``MaxSynBio'' consortium,  C.~Kr{\"u}ger by the German Research Foundation (DFG Priority Programme SPP1726).

\bibliography{manu/Swimmer-Jin.bib}

\newpage
\onecolumngrid
\appendix

\setcounter{equation}{0}
\setcounter{figure}{0}
\renewcommand\thefigure{S\arabic{figure}}
\renewcommand\theequation{S\arabic{equation}}

\section*{Supporting Information}
\subsection*{Summary}
In this document, we provide background information that would exceed the scope of the main article.
\begin{itemize}
 \item Section~S\ref{sec:gradientcalc} provides a detailed derivation of the concentration profile in eqn. 1 of the main text.
 \item Section~S\ref{sec:microfluidics} contains a full image of the branch choice cell, a close-up of which is used in fig. 4 of the main text.
 \item Section~S\ref{sec:conc-effect} provides additional data on the effect of surfactant concentration on the branch choice experiment
 \item Section~S\ref{sec:size-effect} provides additional data on the effect of droplet size on the branch choice experiment
 \item Cover images and captions of supporting movies are provided in the end.
\end{itemize}

\subsection{Derivation of the concentration gradient and the average correlation}\label{sec:gradientcalc}
We approximate the initial distribution of surfactant filled micelles after a swimmer has passed a junction with a two-step function, as sketched in Fig.~5(b):
\begin{equation*}
  c(x, t=0) = 
  \begin{cases}
    c_0 & -l<x<l \\ 
    0 & \text{otherwise}
  \end{cases}
\end{equation*}
We map the junction onto a straight line,  $-L <x < L$, with $L=2l$, and periodic, i.e. reflecting boundaries at $x=\pm L$.
We assume that the solubilized oil is co-moving with the surfactant micelles, and will use a diffusion coefficient valid for surfactant micelles, $D_\text{f}$.

Diffusion from a step profile is a textbook problem, we follow Crank's derivation \cite{crank1979mathematics}: 

\begin{align}
  c(x,t) &= \frac{c_0}{2} \sum_{n=-\infty}^{\infty} \left(\erf \frac{2nL+l-x}{\sqrt{4D_\text{f}t}} - \erf \frac{2nL-l-x}{\sqrt{4D_\text{f}t}}\right) \nonumber \\
   & \stackrel{L=2l}{=}  \frac{c_0}{2} \sum_{n=-\infty}^{\infty} \left(\erf \frac{(4n+1)l-x}{\sqrt{4D_\text{f}t}} - \erf \frac{(4n-1)l-x}{\sqrt{4D_\text{f}t}} \right)
  \label{eqns:c-erf}
\end{align}

The resulting concentration gradient is:
\begin{equation}
  \frac{\partial c(x,t)}{\partial x} = -\frac{c_0}{2\sqrt{\pi D_\text{f}}} \frac{1}{\sqrt{t}} \sum_{n=-\infty}^{\infty} \left(\exp {\frac{-(4nl+l-x)^2}{4D_\text{f}t}} -\exp {\frac{-(4nl-l-x)^2}{4D_\text{f}t}}\right) 
  \label{eqns:dc}
\end{equation}

Evaluating this expression for  $x = l$ yields:
\begin{multline}
  \frac{\partial c(x,t)}{\partial x} \biggr \rvert_{x=l} 
  = \frac{c_0}{2 \sqrt{\pi D_\text{f}}} \frac{1}{\sqrt{t}} \,
  \left[ -1 +\exp\left(\frac{l^2}{D_\text{f}t}\right) \right. \\
   -\exp\left(-\frac{4l^2}{D_\text{f}t}\right) +\exp\left(-\frac{l^2}{D_\text{f}t}\right)
   \left. + \exp\left(-\frac{9l^2}{D_\text{f}t}\right) -\exp\left(-\frac{4l^2}{D_\text{f}t}\right) \dots\right]
  \label{eqns:dc-trun}
\end{multline}
We have truncated the series of infinite reflections at the periodic boundary to the first reflection ($n = 0,\, 1,\, -1$), resulting in the six terms shown above.

The average correlation between the choices of consecutive swimmers is
\begin{align*}
  \langle\mathcal{C}\rangle &= -1 \cdot \mathcal{P} + 1 \cdot (1-\mathcal{P}) =  1-2\mathcal{P} \\ 
  &= \frac{ \exp ( 2\xi ) +  \exp ( -2\xi) -2}{ \exp ( 2\xi ) - \exp ( -2\xi) } \\
  &= \frac{ ( \exp ( \xi ) -  \exp ( -\xi) ) ^2}{( \exp ( \xi ) + \exp ( -\xi ) )  ( \exp ( \xi ) - \exp ( -\xi ) ) } 
  = \tanh (\xi )
\end{align*} 

And with Eq.~\eqref{eqns:dc-trun} and $c_0 = \beta / (vw^2)$, we get
\begin{equation}
  \xi = -\frac{\kappa \cdot d}{2D} \partial_x c =- \frac{a}{\sqrt{t}}(2 \exp (-b/t) -1 - 2 \exp (-4b/t) + \exp (-9b/t) )
  \label{eqns:xi-fit}
\end{equation}
with two parameters, a pre-factor $a$ and a time constant $b$:
\begin{align}
a &= \frac{\kappa}{D} \cdot\frac{\beta \,d}{4vw^2 \sqrt{\pi D_\text{f}}}, & b =\frac{l^2}{D_\text{f}}.
\label{eqns:ab}
\end{align}

\subsection{Design of the branch choice device}\label{sec:microfluidics}
\begin{figure}[!htb]
  \centering
  \includegraphics[width=.8\textwidth]{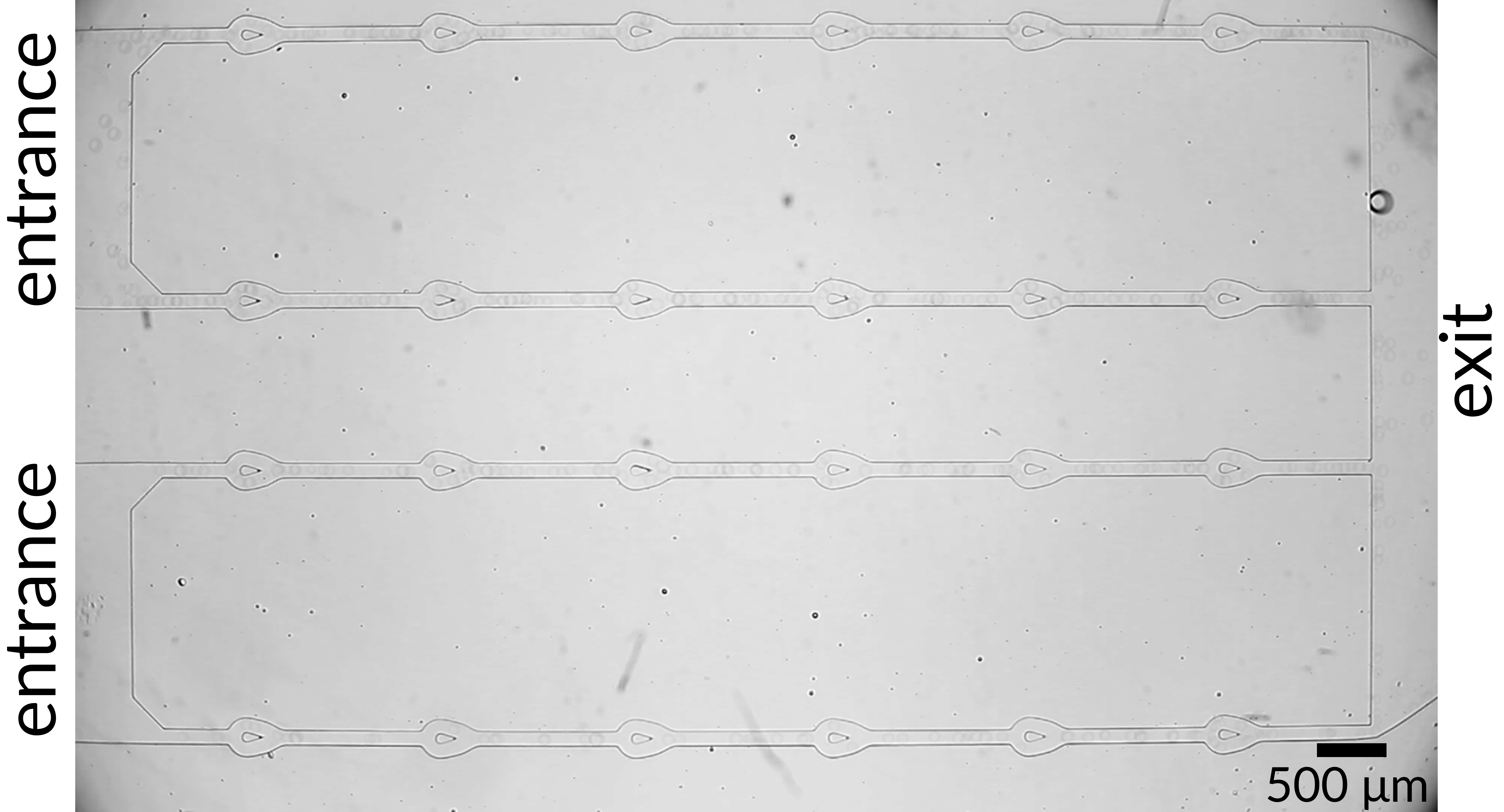}
  \caption{Microfluidic device for the branch-choice experiment.}
  \label{figs:device}
\end{figure}

In order to to observe branch choice incidents in numbers adequate for statistical analysis, we used a multi-pillar device with 4 parallel channels of 6 junctions each. For illustration purposes, we provide a still image of the empty device in fig.~\ref{figs:device}. The channel width is \SI{100}{\um}.

\subsection{The effect of surfactant concentration on the branch choice experiment}\label{sec:conc-effect}
\begin{figure}[!htb]
  \centering
  \includegraphics[width=.9\textwidth]{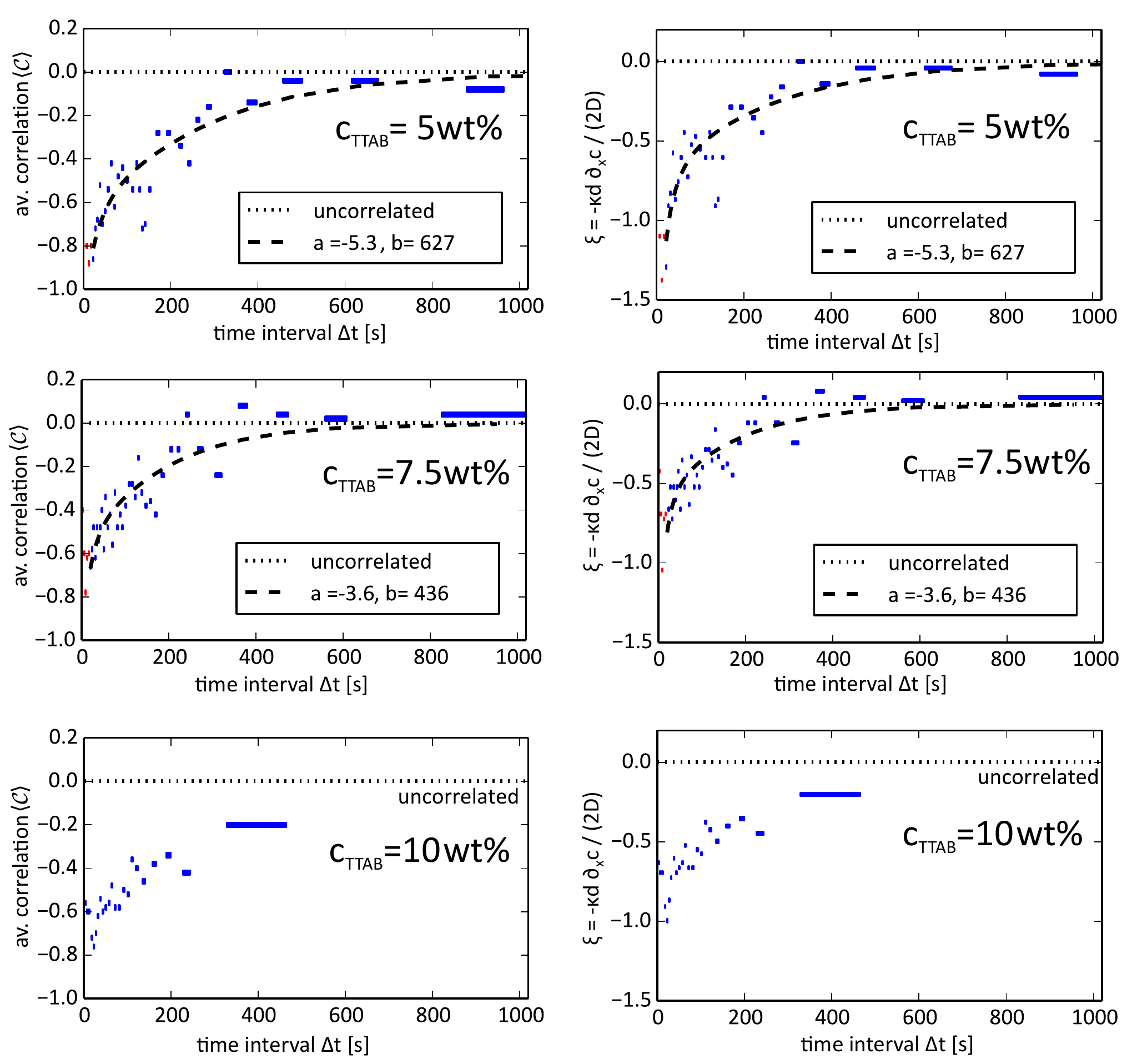}
  \caption{Correlation $\langle\mathcal{C}\rangle$ and $\xi$ vs. the time interval $\Delta t$ at variable $c_\text{TTAB}$. The fit of $\xi$ to our analytical model (Eq.~\eqref{eqns:xi-fit}) and the corresponding function for $\langle\mathcal{C}\rangle$ are drawn in black dashes. As our model is not applicable for small $\Delta t$, the data points marked in red have not been used in the parameter fit. 
  Data were binned by $\Delta t$ using $100$ events/bin and then averaged (the corresponding  $\Delta t$ range is marked by the bar width). By fitting $\xi$ using Eq.~\eqref{eqns:xi-fit}, we obtain for $c_\text{TTAB}=5$\,wt\% the parameters $a=-5.3\pm 0.3$ and $b=(6.3\pm 2.5)\times 10^2\si{\second}$, and for $c_\text{TTAB}=7.5$\,wt\% the parameters $a=-3.6\pm 0.2$ and $b=(4.3\pm 1.2)\times 10^2\si{\second}$.
  We provide no fit for $c_\text{TTAB}=10$\,wt\% due to insufficient statistics in the long time limit.}
  \label{figs:ac-conc}
\end{figure}

We repeated the branch choice experiment under different TTAB concentrations $c_\text{TTAB}$ using $5\,\text{wt}\%$, $7.5\,\text{wt}\%$ (data presented in the main text), $10\,\text{wt}\%$. The range of concentrations accessible in these experiments was limited by the minimum concentration necessary for propulsion and the short droplet lifetime at higher concentrations. Qualitatively, an increase in $c_\text{TTAB}$ should decrease the relative fraction of filled micelles and weaken the observed anticorrelation. However, changing  $c_\text{TTAB}$ affects the experiment in multiple ways:
\begin{itemize}
  \item Since both velocity $v$ and dissolving rate $\beta$ will increase with  $c_\text{TTAB}$, we don't expect large changes in the concentration of dissolved oil, $c_0$.
  \item Our propulsion model~\cite{herminghaus2014interfacial} is based on the assumption that the local surfactant density on the droplet interface is coupled to the density of empty micelles in the vicinity and that oil filled micelles cannot be used to replenish the interface. At the moment, we have no independent data on the kinetics of oil molecule transport between droplet and micelles and in what manner partially filled micelles act as chemorepellents, such that we cannot make quantitative predictions about the relation between oil concentration $c_0$, the auto-chemotactic coupling $\kappa$ and $c_\text{TTAB}$. 
  \item The diffusion coefficient of micelles $D_f$ is dependent on micelle size and viscosity. Assuming a Stokes-Einstein relation, it should decrease with $c_\text{TTAB}$, however, according to literature~\cite{candau1984new} the effect is rather weak in the accessible range of $c_\text{TTAB}$.
\end{itemize}
For all experiments, we first obtained the average correlation $\langle\mathcal{C}\rangle$ by binning $100$ events, then calculated $ \xi = \arctanh (\langle\mathcal{C}\rangle)$. We fitted $\xi$ to the analytical model (Eq.~\eqref{eqns:xi-fit}) to obtain the parameters $a$ and $b$ and provide plots of $\xi$ and $\langle\mathcal{C}\rangle$ vs. time for the 3 data sets. We provide no fit for the 10\,wt\% data set, as the decorrelation timescale exceeds the lifetime of the droplets, which led to insufficient statistics for the long time limit.

Comparing the 5\,wt\% and 7.5\,wt\% data sets, we observe a reduced anti-correlation. 

\subsection{Branch choice experiment with smaller droplets}\label{sec:size-effect}
Another way of varying the experimental parameters without having to change the microfluidic chip design was to use smaller droplets of \SIrange{50}{70}{\um} in diameter in a channel of width \SI{100}{\um} and height \SI{110}{\um}. There are two effects influencing the branch choice behavior. $c_0$ is reduced, as a smaller droplet deposits less oil, which reduces the gradient force. We also expect an influence on the Brownian process characterized by the diffusion coefficient $D$.

The diffusion coefficient $D$ of an active droplet is not well defined in this situation. The effective diffusion coefficient for active Brownian particles as defined in~\cite{howse2007self} is $D_\text{eff}=D_\text{Br}+\frac14v^2\tau_R$ for a swimmer propelling along its polar axis at a speed $v$ with a rotational diffusion time $\tau_R$ and a passive Brownian translational diffusion $D_\text{Br}$. This expression is valid in the time limit of $t\gg \tau_R$, in which case the trajectory resembles a random walk. In our 1D mapping, which is applicable if the droplet is the same size as the branch junction and forward motion is suppressed at the junction by the pillar, i.e. $v=0$, $D=D_\text{passive}>D_\text{Br}$, as noted in the main manuscript. However, if the droplet size is smaller than the junction, the problem is not purely one-dimensional any more, and the Brownian process will also depend on $v$ and $\tau_R$, but not to the limit of  $D=D_\text{eff}$ since $t<\tau_R$ -- the propulsion is close to ballistic on this timescale. Additionally, since the propulsion mechanism is based on Marangoni flows in the droplet interface, we expect $\tau_R$ to depend strongly on the confinement conditions. Because of the additional rotational diffusion contribution, we can assume $D$ to increase.

\begin{figure}[!htb]
  \centering
  \includegraphics[width=.9\textwidth]{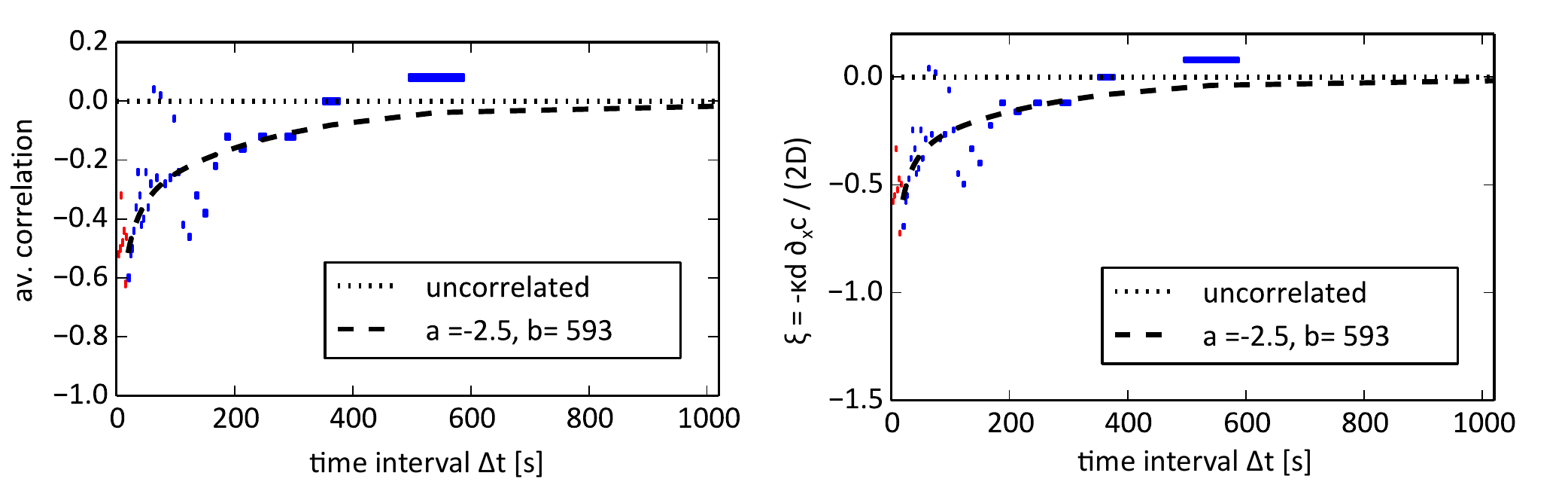}
  \caption{Branch choice experiment with \SI{70}{\um} (diameter) droplets in $5\,\text{wt}\%$ TTAB.
  Data were binned by $\Delta t$ using $100$ events/bin and then averaged (the corresponding  $\Delta t$ range is marked by the bar width). Fitting $\xi$ to Eq.~\ref{eqns:xi-fit} resulted in the parameters $a=-2.5\pm 0.2$ and $b=(5.9\pm 4.7)\times 10^2\si{\second}$.
}
  \label{figs:ac-5wt-70}
\end{figure}

From the theoretical model, we again expect a weaker anti-correlation or decrease in $a$ for the smaller droplets, with a decay time (as reflected by $b$) comparable to the experiments with larger droplets (the micellar dynamics and junction size remain unchanged). This is not inconsistent with the correlation data in Figure \ref{figs:ac-5wt-70}.

\newpage
\section*{Supporting movies}
\begin{figure}[!htb]
  \centering
  \includegraphics[width=.5\textwidth]{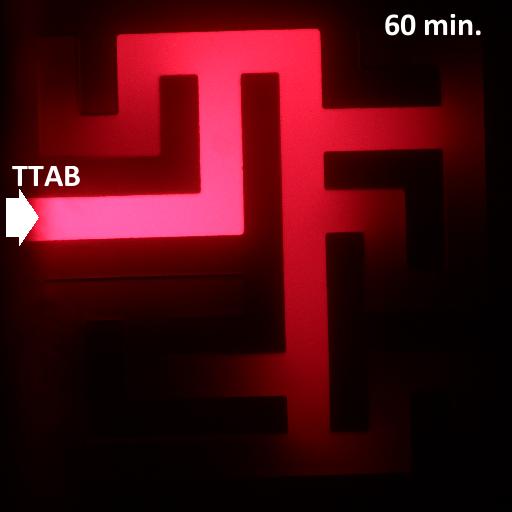}
  \caption{Surfactant spreading through a maze imaged via the fluorescent Nile Red dye.  Experimental duration \SI{60}{\minute}, field of view $2.6 \times 2.6$ \si{\mm}. See Fig.~2(a).}
  \label{mov:spreading}
\end{figure}

\begin{figure}[!htb]
  \centering
  \includegraphics[width=.5\textwidth]{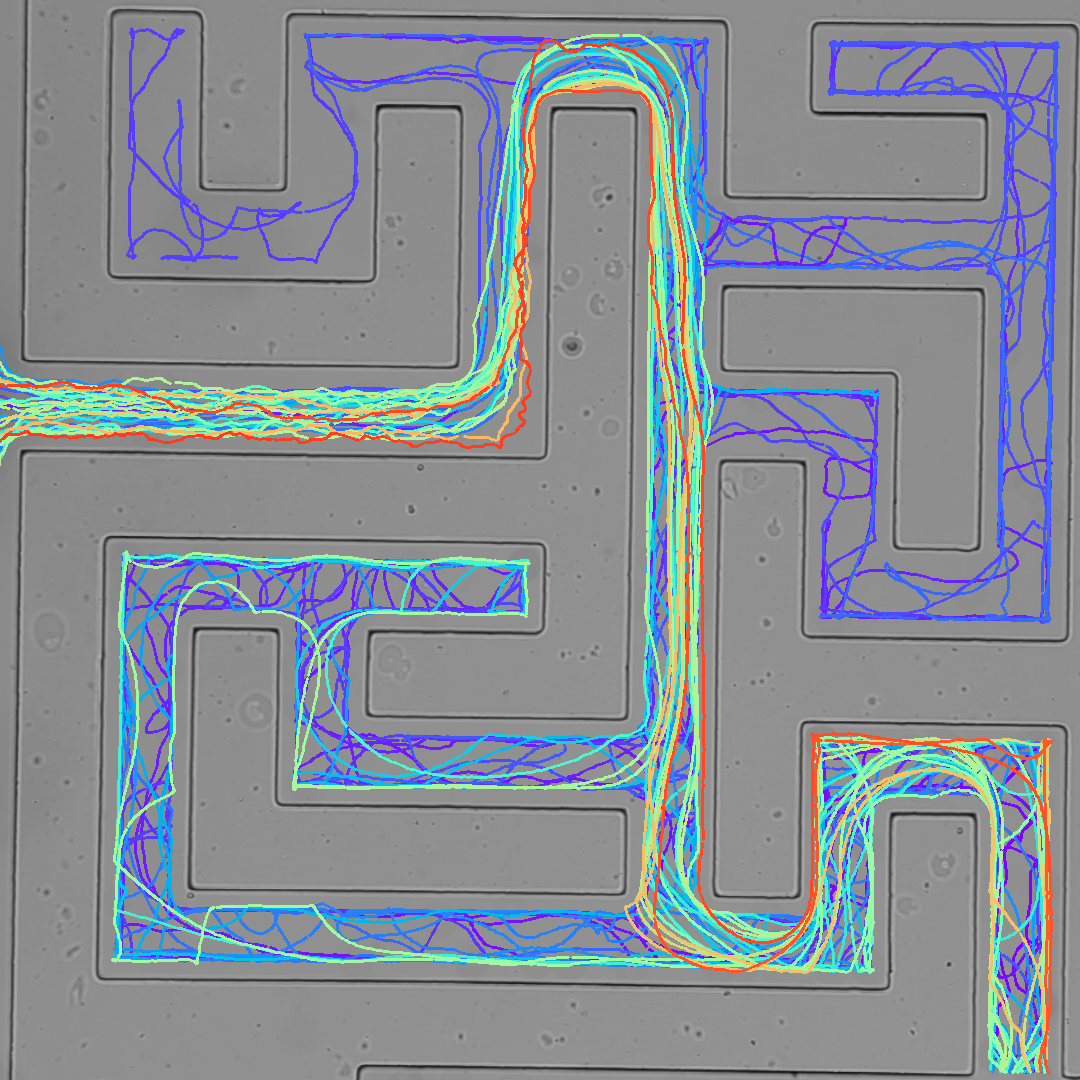}
  \caption{Droplet swimmers (diameter \SI{100}{\um}) are guided through a maze by a TTAB surfactant gradient building up over time. The maze is pre-filled with a $5 \text{wt}\%$ TTAB solution at room temperature to ensure droplet motility.  Experimental duration \SI{83}{\minute}, field of view $2.3 \times 2.3$ \si{\mm}. See Fig.~2(c).}
  \label{mov:maze}
\end{figure}

\begin{figure}[!htb]
  \centering
  \includegraphics[width=.5\textwidth]{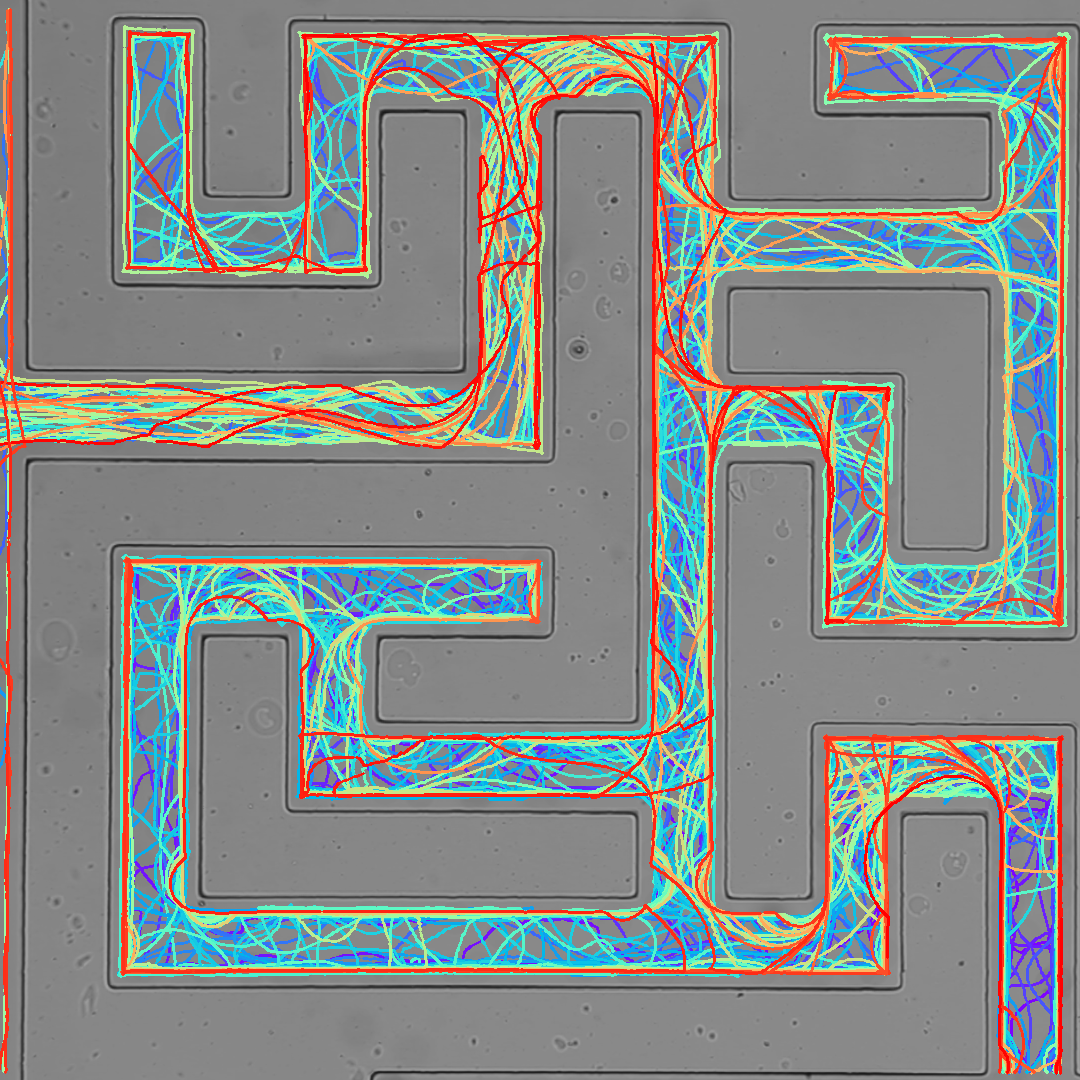}
  \caption{Droplet swimmers (diameter \SI{100}{\um}) exploring a maze pre-filled with an isotropic $5 \text{wt}\%$ TTAB solution at room temperature.  Experimental duration \SI{92}{\minute}, field of view $2.3 \times 2.3$ \si{\mm}. See Fig.~2(d).}
  \label{mov:maze-control}
\end{figure}

\begin{figure}[!htb]
  \centering
  \includegraphics[width=.5\textwidth]{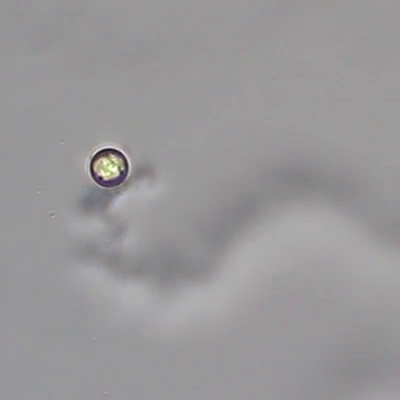}
  \caption{A self-propelling droplet swimmer (size \SI{50}{\um}) leaves a trail of oil-filled micelles, visualized by phase contrast microscopy. To increase the droplet solubilization rate, the experiment was conducted at high TTAB concentration ($25 \text{wt}\%$) and temperature (\SI{37}{\degreeCelsius}). Experimental duration \SI{96}{\second}, field of view $0.36 \times 0.36$ \si{\mm}. See Fig.~3(a).}
  \label{mov:trail}
\end{figure}

\begin{figure}[!htb]
  \centering
  \includegraphics[width=.5\textwidth]{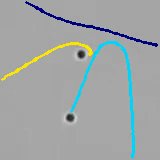}
  \caption{Three droplet swimmers (size \SI{50}{\um}) being repelled from each other's trails. Experimental duration \SI{50}{\second}, field of view $0.94 \times 0.94$ \si{\mm}. See Fig.~3(b).}
  \label{mov:free-drop}
\end{figure}

\begin{figure}
  \centering
  \includegraphics[width=.8\textwidth]{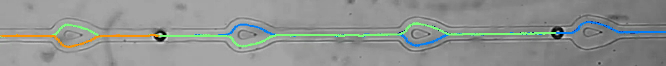}
  \caption{Droplets (size \SI{100}{\um}) navigating bifurcating channels (width \SI{100}{\um}). The branch choices of consecutive droplets are predominantly anticorrelated due to negative autochemotaxis. Experimental duration \SI{7}{\minute} and \SI{28}{\second}, field of view $5.9 \times 0.58$ \si{\mm}. See Fig.~4.}
  \label{mov:branch}
\end{figure}

\end{document}